\documentclass[lettersize,journal]{IEEEtran}
\usepackage{amsmath,amsfonts}
\usepackage{algorithmic}
\usepackage{algorithm}
\usepackage{array}
\usepackage[caption=false,font=normalsize,labelfont=sf,textfont=sf]{subfig}
\usepackage{textcomp}
\usepackage{stfloats}
\usepackage{url}
\usepackage{verbatim}
\usepackage{graphicx}
\usepackage{cite}
\usepackage{xcolor} 
\definecolor{DarkBlue}{RGB}{8,81,156}
\definecolor{CBBlue}{RGB}{49,130,189} 
\definecolor{CBRed}{RGB}{227,26,28}   

\usepackage{siunitx}
\usepackage{eurosym}
\usepackage{pgfplots}
\pgfplotsset{compat=1.18}

\begin{document}

\title{Bidding Aggregated Flexibility in European Electricity Auctions}

\author{Gabriel~Ellemund, Thomas~Hübner, Quentin~Lété, Stefano~Bracco, Matteo~Fresia,
Gabriela~Hug
\thanks{This work was supported by the Swiss Federal Office of Energy’s SWEET program through the PATHFNDR project [Grant Number SI/502259].}
\thanks{G. Ellemund, T. Hübner, and G. Hug are with the Power Systems Laboratory, ETH Zurich, 8092 Zurich,
Switzerland.}%
\thanks{Q. Lété is with CORE/LIDAM, Université catholique de Louvain, Louvain-la-Neuve, Belgium.}
\thanks{S. Bracco and M. Fresia are with the Department of Electrical, Electronics, and Telecommunication Engineering and Naval Architecture, University of Genoa, 16145 Genoa, Italy.}
\thanks{Correspondence should be addressed to Thomas Hübner (e-mail: thuebner@ethz.ch).}
\thanks{The authors would like to thank Carlo Tajoli, Katharina Kaiser, and Adrien Mellot for their valuable comments and discussions related to this work.}
}



\maketitle

\begin{abstract}
Bidding flexibility in day-ahead and intraday auctions would enable decentralized flexible resources, such as electric vehicles and heat pumps, to efficiently align their consumption with the intermittent generation of renewable energy. However, because these resources are individually too small to participate in those auctions directly, an aggregator (e.g., a utility) must act on their behalf.  This requires aggregating many decentralized resources, which is a computationally challenging task. In this paper, we propose a computationally efficient and highly accurate method that is readily applicable to European day-ahead and intraday auctions. Distinct from existing methods, we aggregate only economically relevant power profiles, identified through price forecasts. The resulting flexibility is then conveyed to the market operator via exclusive groups of block bids. We evaluate our method for a utility serving the Swiss town of Losone, where flexibility from multiple heat pumps distributed across the grid must be aggregated and bid in the Swiss day-ahead auction. Results show that our method aggregates accurately, achieving 98\,\% of the theoretically possible cost savings. This aggregation accuracy remains stable even as the number of heat pumps increases, while computation time grows only linearly, demonstrating strong scalability.
\end{abstract}


\section{Introduction} \label{sec: introduction}
\IEEEPARstart{A}{s} dispatchable fossil-fuel-based generators are gradually replaced by variable renewable energy sources (RES), coordinating supply and demand becomes increasingly more challenging. This makes demand-side flexibility crucial for maintaining the balance in the electric energy system, with power markets serving as the primary coordination mechanism \cite{cramton2017electricity}.
By matching supply offers from generators with demand bids from consumers, power markets determine the market-clearing price for each time interval. These prices serve as scarcity signals, guiding both generation and consumption toward system needs: flexible generators produce when prices are high, while flexible consumers shift demand to periods of low prices and abundant renewable energy supply \cite{hirth2013market}.

The electrification of heating and transport introduces substantial flexible demand into the energy system. Load shifting of heat pumps (HPs) and electric vehicles (EVs) can alleviate temporal mismatches between supply and demand, supporting RES integration while reducing curtailment and dependence on costly storage solutions~\cite{mellot2025exploratory}. Yet, the flexibility of these small-scale resources remains largely unexploited, as they are typically operated individually and without optimization. To unlock this potential, aggregators must offer the aggregated flexibility of individual resources on the market~\cite{plaum2022aggregated}.


For this purpose, the aggregator needs to compute the aggregated flexibility of the individual resources. Mathematically, the flexibility of a single resource can be represented as a set in a multi-dimensional power space, encompassing all feasible operating states of the resource over a given time horizon while accounting for intertemporal constraints such as energy limits~\cite{zhao2016extracting}. Aggregating multiple resources then amounts to computing the Minkowski sum of their individual feasible sets - an NP-hard problem that quickly becomes intractable for large populations and long time horizons \cite{tiwary2008hardness}. A growing body of literature addresses this problem by developing inner \cite{taha2024efficient,müller2015aggregation,müller2019aggregation,ozturk2024alleviating,nazir2018inner} and outer \cite{barot2017concise,zhao2017geometric,barot2016outer} approximation methods, with the goal of finding approaches that are both accurate and computationally tractable \cite{ozturk2022aggregation}. However, these methods do not discuss how to use the aggregate feasible sets in markets, limiting their practical applicability in systems operation. 

Market participation requires that bids are expressed in a format defined by the market operator. In European day-ahead and intraday auctions, market participants can choose between different bid formats, which allow them to express their willingness to buy or sell electricity, based on their preferences and constraints \cite{herrero2020evolving}. 
In recent years, exclusive groups of block bids (also known as XOR package bids) have been identified as a promising bid format, especially for those resources with significant flexibility~\cite{herrero2020evolving,karasavvidis2024optimal,Hubner2024, hubner2025package}. 

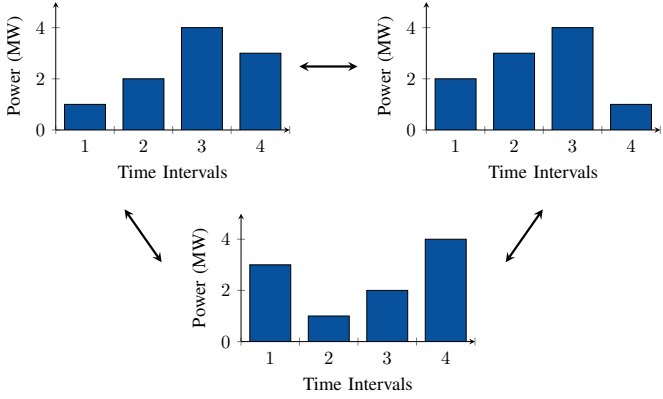
\begin{figure}[t]
    \centering
    \resizebox{1\columnwidth}{!}{%
    \begin{tikzpicture}[every node/.append style={font=\normalsize}, >=stealth]
        \begin{axis}[
            ybar interval=0.7,
            xlabel={Time Intervals},
            ylabel={Power (MW)},
            ymin=0, ymax=5,
            width=6cm, height=4cm,
            enlarge x limits=0.05,
            xtick={1,2,3,4,5},
            xtick align=outside,
            axis x line=bottom,
            axis y line=left,
            ytick pos=left,
            grid=none
        ]
            \addplot[fill=DarkBlue] coordinates {(1,1) (2,2) (3,4) (4,3) (5,2)};
        \end{axis}

        \begin{scope}[xshift=7cm]
        \begin{axis}[
            ybar interval=0.7,
            xlabel={Time Intervals},
            ylabel={Power (MW)},
            ymin=0, ymax=5,
            width=6cm, height=4cm,
            enlarge x limits=0.05,
            xtick={1,2,3,4,5},
            xtick align=outside,
            axis x line=bottom,
            axis y line=left,
            ytick pos=left,
            grid=none
        ]
            \addplot[fill=DarkBlue] coordinates {(1,2) (2,3) (3,4) (4,1) (5,2)};
        \end{axis}
        \end{scope}

        \begin{scope}[xshift=3.5cm, yshift=-4cm]
        \begin{axis}[
            ybar interval=0.7,
            xlabel={Time Intervals},
            ylabel={Power (MW)},
            ymin=0, ymax=5,
            width=6cm, height=4cm,
            enlarge x limits=0.05,
            xtick={1,2,3,4,5},
            xtick align=outside,
            axis x line=bottom,
            axis y line=left,
            ytick pos=left,
            grid=none
        ]
            \addplot[fill=DarkBlue] coordinates {(1,3) (2,1) (3,2) (4,4) (5,2)};
        \end{axis}
        \end{scope}

        \draw[<->, very thick] (4.6, 1.2) -- (5.7, 1.2);  
        \draw[<->, very thick] (9.2, -1.5) -- (8.5, -2.5);  
        \draw[<->, very thick] (2.0, -2.5) -- (1.3, -1.5);  

    \end{tikzpicture}
    }
    \caption[Exclusive group of three block bids]{Exclusive group of three block bids.}
    \label{fig:exclusive_groups}
\end{figure}

Traditionally, bids in power markets were submitted for different time periods separately~\cite{cramton2017electricity}. To account for intertemporal constraints, block bids were introduced, enabling agents to bid on entire power profiles spanning multiple periods~\cite{herrero2020evolving}. By grouping several block bids into an exclusive group (EG), the market-clearing algorithm is constrained to accept at most one of these bids~\cite{Euphemia2024}.
Thus, an EG allows an agent to communicate $B$ possible schedules (power profiles) to the auctioneer while guaranteeing that at most one is activated. Fig.~\ref{fig:exclusive_groups} illustrates the concept: a shiftable load conveys its flexibility as an exclusive group of three block bids. The auctioneer can then select the schedule that best aligns with renewable generation by matching it against sell offers from renewable sources.

The use of EGs in European day-ahead auctions is illustrated in Table~\ref{tab:bid_data}, which reports the total number of submitted EGs and the median number of bids per EG for the bidding zones Austria (AT), Belgium (BE), France (FR), Germany-Luxembourg (DE/LX), Great Britain (GB), the Netherlands (NL), and Switzerland (CH) in 2024. The data provide three key insights:
\begin{itemize}
    \item EGs are still used predominantly on the supply side, although some agents already employ them to express demand-side flexibility.  
    \item Usage varies significantly across countries: while agents in some bidding zones rely heavily on EGs, in others they are scarcely applied.  
    \item The maximum of 24 block bids per EG in the coupled European day-ahead auction~\cite{EPEXSPOT2025, NordPool2025} is often not fully used. 
\end{itemize}
This indicates that, although market participants have already recognized the usefulness of EGs, their widespread adoption, particularly for communicating demand-side flexibility, remains largely unrealized. In particular, the frequent underuse of the 24-block limit suggests that less flexibility is being communicated than is likely available.

\begin{table}[b]
\centering
\caption{Bid data of submitted exclusive groups in 2024.}
\begin{tabular}{ c c c c c }
\hline
Bidding & Demand- & Median buy & Supply- & Median sell \\ 
zone & side EGs & bids/group & side EGs & bids/group \\ \hline
AT     & 52   & 1  & 1     & 2  \\
BE     & 448  & 24 & 2,483 & 21 \\
FR     & 2,259& 17 & 11,260& 22 \\
DE/LX  & 2,101& 13 & 12,589& 15 \\
GB     & 711  & 8  & 12,083& 24 \\
NL     & 298  & 12 & 5,923 & 24 \\
CH     & 19   & 12 & 272   & 20 \\ \hline
\end{tabular}\\
\vspace*{2pt}
{\textit{Note.} The bid data were purchased from the power exchange EPEX SPOT.}
\label{tab:bid_data}
\end{table}

The limit of 24 block bids per EG is introduced by the market operator to ensure computational tractability of the market-clearing algorithm, shifting the task of pre-selecting suitable power profiles to market participants. This limitation introduces a \emph{missing bids problem}, where participants are not able to bid on all of their possible power profiles~\cite{goetzendorff2015compact}. 
Recent studies have proposed bid selection algorithms to tackle this problem. However, existing studies focus exclusively on single assets such as hydropower plants \cite{aminabadi2025optimizing}, flexible loads \cite{karasavvidis2024optimal,hubner2025package}, thermal generators \cite{hubner2025package}, or battery energy storage systems \cite{hubner2025package}, rather than on aggregated portfolios.

Our work bridges this gap with respect to flexibility aggregation and market bidding by introducing a method directly applicable to European day-ahead and intraday auctions. Unlike the prevailing approaches in the flexibility aggregation literature, which aim to approximate the entire aggregated feasible region as closely as possible \cite{ozturk2022aggregation}, we concentrate on identifying the subset that is economically relevant and conveying it to the market operator through EGs of block bids. This economically relevant region is identified by leveraging day-ahead and intraday price forecasts - readily available in practice \cite{lago2021forecasting} and widely used in the bidding literature to guide optimal bid construction \cite{hubner2025package}.

Note that our method represents an application of the generic bidding framework introduced in~\cite{hubner2025package} to the setting of flexibility aggregators. While that work provides the theoretical foundation, we focus on its concrete implementation for flexibility aggregators bidding in day-ahead and intraday auctions.

To evaluate the performance of our proposed \textit{market-applicable flexibility aggregation method}, we apply it to a case study of a utility serving the Swiss town of Losone. In the considered setting, flexibility is provided by HPs distributed across the grid and is communicated to the Swiss day-ahead auction. We consider both the perspective of an \textit{unbundled utility} (supplying customers without managing the grid) and an \textit{integrated utility} (supplying customers while also managing the grid). This distinction reflects the Swiss integrated retail market, where local distribution system operators (DSOs) also act as electricity suppliers to households. By incorporating the distribution grid into our framework, we assess how aggregated flexibility can be harnessed not only in a system-friendly manner to reduce electricity procurement costs, but also in a grid-friendly manner to mitigate local network congestion.

The remainder of this paper is structured as follows. In Section \ref{sec: methodology}, the methodology of the aggregation and bid selection algorithm is described. Section \ref{sec: case_study} focuses on the Losone case study, gives the mathematical formulation of the optimization model, and presents the results. Finally, Section \ref{sec: conclusion} concludes the paper with a summary.

\section{Methodology}\label{sec: methodology}
We consider $R$ flexible resources, indexed by $r \in \mathcal{R}~=~\{1,\dots,R\}$, each operating over $T$ time periods, indexed by $t \in \mathcal{T} = \{1,\dots,T\}$.  
Each resource $r$ has a feasible set $\mathcal{X}_r \subset \mathbb{R}^T$ representing all technically possible consumption (positive) or generation (negative) profiles, and a valuation function $v_r: \mathcal{X}_r \to \mathbb{R}$ representing the cost or benefit associated with a profile $x_r$.

\begin{figure}[h!]
    \centering
    \begin{tikzpicture}[scale=0.9, every node/.append style={font=\small}, >=stealth]

        \node[draw, minimum width=3cm, minimum height=0.75cm] (HPs) at (0,0) {Flexible resources};
        \node[draw, minimum width=3cm, minimum height=0.75cm] (Aggregator) at (0,2) {Aggregator};
        \node[draw, minimum width=3cm, minimum height=0.75cm] (Market) at (0,4) {Auction};

        \draw[->, thick] ([xshift=-1.0cm]HPs.north) -- ([xshift=-1.0cm]Aggregator.south);
        \draw[->, thick] ([xshift=1.0cm]Aggregator.south) -- ([xshift=1.0cm]HPs.north);

        \draw[->, thick] ([xshift=-1.0cm]Aggregator.north) -- ([xshift=-1.0cm]Market.south);
        \draw[->, thick] ([xshift=1.0cm]Market.south) -- ([xshift=1.0cm]Aggregator.north);

        \node[anchor=east] at ([xshift=-1.0cm,yshift=0.6cm]HPs.north) {Aggregation};
        \node[anchor=west] at ([xshift=1.0cm,yshift=0.6cm]HPs.north) {Disaggregation};

        \node[anchor=east] at ([xshift=-1.0cm,yshift=0.6cm]Aggregator.north) {Market bidding};
        \node[anchor=west] at ([xshift=1.0cm,yshift=0.6cm]Aggregator.north) {Market clearing};

    \end{tikzpicture}
    \caption{Aggregation steps between flexible resources, aggregator, and auction.}
    \label{fig:market_structure}
\end{figure}
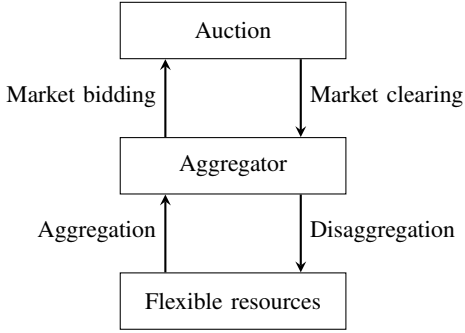

The \emph{aggregate feasible set} $\mathcal{X}_{\mathrm{agg}}$ contains all power profiles that the aggregator can collectively implement. Mathematically, it corresponds to the Minkowski sum of the individual feasible sets:
\begin{align*}
\mathcal{X}_{\mathrm{agg}} := \; \bigoplus_{r=1}^R \mathcal{X}_r  = \; \left\{ x \in \mathbb{R}^T \ \middle| \ x = \sum_{r=1}^R x_r, \; x_r \in \mathcal{X}_r \right\}.
\end{align*}
Similarly, the aggregated valuation function representing cost/benefit is given by 
$$
v_{\mathrm{agg}}(x) := \; \sum_{r=1}^R v_r(x_r).
$$
In the following, we explain how an aggregator can bid approximations of 
$\mathcal{X}_{\mathrm{agg}}$ and $v_{\mathrm{agg}}$ into the European day-ahead or intraday 
auctions via an exclusive group of block bids. We then outline how, based on the aggregator's bids, these auctions determine a 
dispatch $\bar{x}^\ast\in\mathcal{X}_{\mathrm{agg}}$ for the aggregator through market clearing. Subsequently, we show how 
this aggregated dispatch can be disaggregated into individual dispatches $\bar{x}^\ast_r$ for 
each resource $r$. This procedure is summarized in Fig.~\ref{fig:market_structure} and explained 
in detail below.

\subsection{Market Bidding - Exclusive Groups of Block Bids}
A block bid is given by $(x_b,p_b)$ where $x_b=(x_{b1}, \ldots, x_{bT})$ is a power profile and $p_b\in\mathbb{R}$ is the associated price.
The acceptance rate of a bid is given by $\alpha_b\in[0,1]$. The auctioneer can reject ($\alpha_b=0$), fully accept ($\alpha_b=1$) or partially accept ($\alpha_b\in(0,1)$) a bid.
Multiple block bids $b=1,\ldots,B$ can be submitted as part of an EG:
\begin{subequations}\label{eq: exclusive group}
\begin{equation}
\mathcal{G} = \{ (x_b, p_b) \mid b = 1,\dots,B \}.
\end{equation}
Submitting an EG ensures that at most one bid is fully accepted, or that multiple bids are partially 
accepted such that their total acceptance rate does not exceed one (constituting a convex combination):
\begin{equation}
    \sum_{b=1}^B \alpha_b \le 1 .
\end{equation}
\end{subequations}

\subsection{Aggregation - Determining $B$ Block Bids}
To determine $B$ block bids $(x_b, p_b)$, we assume that a forecast of the electricity prices 
$\lambda = (\lambda_1, \ldots, \lambda_T)$ is available, 
represented through $S = B$ price scenarios $\lambda_s \in \mathbb{R}^T$.
For each scenario $s$ and each resource $r$, we determine the optimal dispatch given prices $\lambda_s$ by solving the profit-maximization problem
\begin{equation}
\label{eq: aggregation}
\bar{x}_{r,s} \in \arg\max_{x_r \in \mathcal{X}_r} \left[ v_r(x_r) - \langle \lambda_s, x_r \rangle \right],
\end{equation}
where $\langle \lambda_s, x_r \rangle = \sum_{t=1}^T \lambda_{s,t} x_{r,t}$ represents the forecasted revenue or cost, associated with schedule $x_r$ under price scenario $s$. 
Consequently, the aggregated profile for scenario $s$ is
\begin{equation*}
\bar{x}_s = \sum_{r=1}^R \bar{x}_{r,s}, \qquad
p_s = \sum_{r=1}^R v_r(\bar{x}_{r,s}),
\end{equation*}
where $\bar{x}_s$ and $p_s$ denote the aggregated power profile and the bid price under truthful bidding for scenario $s$, respectively.  
The resulting exclusive group $\mathcal{G}$ consisiting of $B=S$ bids is given by
\begin{equation*}
\mathcal{G} = \{ (\bar{x}_s, p_s) \mid s = 1,\dots,S \}.
\end{equation*}
Note that aggregation is relatively simple, as just $R \cdot S$ small and easy-to-solve subproblems~\eqref{eq: aggregation} need to be solved.

\subsection{Market Clearing - Dispatch of the Aggregator}\label{paragraph: market clearing}
The algorithm EUPHEMIA used to clear the coupled European day-ahead and intraday auctions determines (approximately) a Walrasian equilibrium, where total supply matches total demand and no participant has an incentive to deviate from the outcome~\cite{bichler2021walrasian,hubner2025approximate}.
This means that only the optimal bid of each agent is accepted. 
That is, the market-clearing algorithm always accepts the most profitable bid $(\bar{x}^*, p^*)$ from $\mathcal{G} \cup {(0,0)}$, or, if multiple bids yield the same profit, a combination of partially accepted bids:
\begin{subequations}\label{eq: bid acceptance}
\begin{align}
(\bar{x}^*, p^*) \in & \; \arg\max_{\alpha} \; \sum_{s \in \mathcal{G}} \alpha_s \big(p_s - \langle \tilde{\lambda}, x_s \rangle \big) \\
& \qquad \text{s.t.} \quad 0 \leq \sum_{s \in \mathcal{G}} \alpha_s \leq 1, \\
& \qquad \qquad 0 \leq \alpha_s \leq 1 \quad \forall s \in \mathcal{G}.
\end{align}
\end{subequations}

\subsection{Disaggregation - Dispatch of the Individual Resources}
Once the auctioneer returns the accepted aggregated profile $\bar{x}^*$ from~\eqref{eq: bid acceptance}, disaggregation to individual resources is straightforward in our framework. For the accepted bid $s^*$ that produced $\bar{x}^*$, we already have the optimal per-resource profiles $\{\bar{x}_{r,s^*}\}_{r=1}^R$ from the aggregation step. The aggregator can therefore instruct each resource $r$ to follow its corresponding $\bar{x}_{r,s^*}$ without any additional optimization, ensuring that all technical constraints are satisfied. Similarly, if bids are only partially accepted, each resource $r$ simply follows the corresponding convex combination of its individual schedules, which preserves feasibility since all $\bar{x}_{r,s}$ were optimal and feasible by construction.

\subsection{Algorithm}
The proposed aggregation and bid determination method is summarized in Algorithm~\ref{alg:agg_bid_selection}.
\begin{algorithm}[H]
\caption{Aggregation and bid determination}
\label{alg:agg_bid_selection}
\begin{algorithmic}[1]
\STATE \textbf{Input:} $S$ price scenarios $\{\lambda_s\}_{s=1}^S, \; \lambda_s \in \mathbb{R}^T$
\FOR{$s = 1,\dots,S$}
    \FOR{$r = 1,\dots,R$}
        \STATE Solve $\bar{x}_{r,s} \gets \arg\max\limits_{x_r \in \mathcal{X}_r} \big[ v_r(x_r) - \langle \lambda_s, x_r \rangle \big]$
    \ENDFOR
    \STATE Aggregate: $\bar{x}_s \gets \sum_{r=1}^R \bar{x}_{r,s}, \quad p_s \gets \sum_{r=1}^R v_r(\bar{x}_{r,s})$
\ENDFOR
\STATE \textbf{Output:} Exclusive group $\mathcal{G} \gets \{ (\bar{x}_s, p_s) \}_{s=1}^S$
\end{algorithmic}
\end{algorithm}

\subsection{Example}
\begin{figure}[htbp]
    \centering

    \begin{tikzpicture}[scale=0.5]
        \draw[->] (-5,0) -- (8,0) node[below] {$x_1$ (kW)};
        \draw[->] (0,-6) -- (0,6) node[left] {$x_2$ (kW)};

        \foreach \x in {-4,-2,2,4} \draw (\x,0.2) -- (\x,-0.2) node[below] {\x};
        \foreach \y in {-4,-2,2,4} \draw (0.2,\y) -- (-0.2,\y) node[left] {\y};

        \draw[dashed,blue] (-2.5,0) -- (-2.5,5.5) -- (3,5.5) -- (5.5,3) --  (5.5,-5.5) -- (3,-5.5) -- cycle;
        \fill[blue!40,opacity=0.2] (-2.5,0) -- (-2.5,5.5) -- (3,5.5) -- (5.5,3) --  (5.5,-5.5) -- (3,-5.5) -- cycle;
        \fill[black!30,opacity=0.5] (-2,0) -- (-2,5) -- (3,5) -- (5,3) --  (5,-5) -- (3,-5) -- cycle;
        \draw[thick,black] (-2,0) -- (-2,5) -- (3,5) -- (5,3) --  (5,-5) -- (3,-5) -- cycle;         
        \draw[dashed,green] (-1.5,0) -- (-1.5,4.5) -- (3,4.5) -- (4.5,3) --  (4.5,-4.5) -- (3,-4.5) -- cycle;        
        \fill[green!40,opacity=0.2] (-1.5,0) -- (-1.5,4.5) -- (3,4.5) -- (4.5,3) --  (4.5,-4.5) -- (3,-4.5) -- cycle;      
       
        \fill[red!40,opacity=0.2] (-2,0) --  (5,3) --  (5,-5) -- cycle;
        \draw[dashed,red] (-2,0) --  (5,3) --  (5,-5) -- cycle;

        \node[fill=red, circle, inner sep=2pt] at (5,-5) {};
        \node[fill=red, circle, inner sep=2pt] at (-2,0) {};
        \node[fill=red, circle, inner sep=2pt] at (5,3) {};
        
    \end{tikzpicture}

    \caption{Approximation of the aggregate feasible set (black) using our method (red) and an inner (green) as well as outer approximation (blue).}
    \label{fig: polytope}
\end{figure}
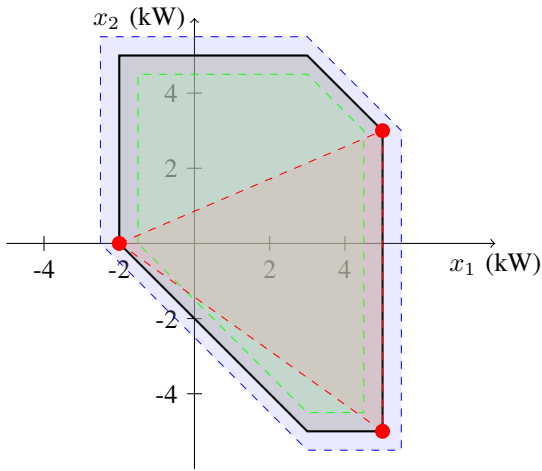

As an illustrative example, consider the aggregate feasible set $\mathcal{X}_{\mathrm{agg}}$ in Fig. \ref{fig: polytope} depicted by the black lines. Here, $x_1$ and $x_2$ denote the power production (negative values) and consumption (positive values) of the aggregated portfolio in time steps 1 and 2, respectively. The aggregated feasible set includes all possible power profiles of the aggregated resources within this two-hour period. Inner approximations (green lines) only cover part of the feasible set and may underestimate the available flexibility, while outer approximations (blue lines) cover the full set but also include infeasible points. On the contrary, our method samples three power profiles (red dots) from $\mathcal{X}_{\mathrm{agg}}$. The resulting EG then spans the red area as any convex combination of block bids in an EG can be accepted, see eq.~\eqref{eq: exclusive group}.

\subsection{Discussion}

Ideally, an aggregator would submit the entire feasible set $\mathcal{X}_{\mathrm{agg}}$ and 
valuation $v_{\mathrm{agg}}(x)$ to the auction. If $\mathcal{X}_{\mathrm{agg}}$ is a polyhedron, 
this could be achieved by bidding every vertex (corner point) as a block bid. In 
Figure~\ref{fig: polytope}, this corresponds to placing 6 block bids. The dispatch obtained 
through market clearing~\eqref{eq: bid acceptance} would then be the aggregator’s optimal dispatch.\footnote{Note that, by inspecting~\eqref{eq: bid acceptance}, it is clear that this optimal dispatch could 
also be obtained if the aggregator had prior knowledge of the market-clearing prices 
$\tilde{\lambda}$.}

In practice, however, the auctioneer imposes a bid limit $B$, requiring the aggregator to 
approximate $\mathcal{X}_{\mathrm{agg}}$, as illustrated in Figure~\ref{fig: polytope} assuming $B=3$. The necessity of such a limit becomes apparent when considering that the optimization 
problem used to clear the coupled European day-ahead auction must determine an economic dispatch 
for the entire continent. By approximating $\mathcal{X}_{\mathrm{agg}}$ using forecasts of the 
prices $\tilde{\lambda}$ - which correspond to the dual variables of the market’s optimization 
problem - we can minimize approximation losses. See~\cite{hubner2025package} for further details on the theoretical background.

\subsection{Network - The Case of an Integrated Utility}\label{paragraph: network}
Note that so far we have assumed the resources $r = 1, \ldots, R$ to be independent. This 
assumption holds for aggregators (\textit{unbundled utilities}) operating in countries with liberalized 
retail markets, such as Germany or the UK. In contrast, in countries like Switzerland without a 
liberalized retail market, \textit{integrated utilities} exist, which must both supply customers 
and manage the distribution grid. In this case, the individual feasible sets $\mathcal{X}_r$ of 
the resources are no longer independent, since the resources are coupled through network 
constraints, i.e., $(x_1, \ldots, x_R) \in \mathcal{F}$. Therefore, for each scenario $s$, all resources are jointly optimized in a single optimization problem:
\begin{align*}
\bar{x}_s \in & \; \arg\max \; \sum_{r=1}^R \big[ v_r(x_r) - \langle \lambda_s, x_r \rangle \big] \\
& \qquad \text{s.t.} \quad x_r \in \mathcal{X}_r, \quad \forall r \in \mathcal{R}, \\
& \qquad \qquad  (x_1, \ldots, x_R) \in \mathcal{F}.
\end{align*}
This approach replaces the $R \cdot S$ independent optimization problems of the unbundled case with $S$ optimal power flow (OPF) problems.
For an explicit formulation of the network constraints $(x_1, \ldots, x_R) \in \mathcal{F}$, we refer the reader to our case study in the distribution grid of Losone, Switzerland presented in Section~\ref{sec: case_study}.

\section{Case Study} \label{sec: case_study}
In this section, the proposed aggregation and bidding method is tested on a realistic case study. We focus on the load-shifting flexibility of HPs to optimally schedule their operation in the day-ahead market (DAM) with the objective of minimizing electricity procurement costs. Due to the thermal inertia of buildings, HPs can shift consumption from periods of high prices to periods of low prices with only minor effects on indoor temperature. This short-term flexibility potential of HPs has been widely studied, e.g., in~\cite{marijanovic2022value,golmohamadi2021optimization}.


We apply our approach to both an unbundled and an integrated utility, representing an aggregator in a liberalized and monopoly-based retail market, respectively. We assume that the aggregator is allowed to directly control the HPs, but must ensure thermal comfort to end-users by keeping the indoor temperature within predefined temperature bounds. 

All optimization models are formulated using JuMP in Julia version 1.11.5 and solved using Gurobi version 11.0.3 with default settings \cite{Gurobi} on a laptop with 32 GB RAM and an AMD Ryzen 7 PRO 6850U processor.

\subsection{Data} \label{subsec: case_study_description}
The case study is calibrated on data from the municipality of Losone, Switzerland. The municipality comprises 2400 buildings, of which 350 are already equipped with HPs (approximately 15\,\%). From this initial configuration, we increase the HP share in increments of 15\,\% up to a maximum penetration of 60\,\%. For this, buildings with HPs are selected randomly, with the condition that once a building is equipped with an HP, it retains it in all subsequent scenarios with higher HP shares. This setup tests both the scalability of our method and shows to what extent the flexibility can be leveraged for peak shaving when the system demand increases, compared to an inflexible case. The number and capacity of HP installations for the different penetration levels are summarized in Table~\ref{tab:hps_installed}.
\begin{table}[h!]
\centering
\caption{heat pump installations in Losone.}
\renewcommand{\arraystretch}{1.1} 
\begin{tabular}{ c c c }
\hline
Installed HPs & Total HP share & Installed HP electrical capacity \\ \hline
350 & 15\,\% & \SI{0.93}{\mega\watt} \\
720 & 30\,\%& \SI{2.00}{\mega\watt} \\
1080  & 45\,\% & \SI{3.00}{\mega\watt} \\
1440  & 60\,\% & \SI{4.00}{\mega\watt} \\ \hline
\end{tabular}
\label{tab:hps_installed}
\end{table}

As the flexibility potential provided by HPs is mainly available in the heating season, we focus on the period from October 1, 2024, to March 31, 2025. The method is applied to the DAM with a time horizon of $T$ = 24 hours and time intervals of $\Delta t$ = 1 hour. The optimization model is solved separately for each day $d$ and scenario $s$. For simplicity, these indices are omitted in the following formulation of the building and network model.

The input data for the case study are retrieved from the sources summarized in Table \ref{tab:input_data}, whereas all fixed input parameters are shown in Table \ref{tab:constants}. 
\begin{table}[h!]
\centering
\caption{Overview of data sources for the case study.}
\renewcommand{\arraystretch}{1.00} 
\begin{tabular}{ l l }
\hline
\textbf{Input data} & \textbf{Data source} \\
\hline
Outdoor temperatures & Open Meteo~\cite{OpenMeteo} \\
Building parameters & Wen et al.~\cite{wen2025quantifying} \\
Load & ENTSO-E transparency platform~\cite{ENTSOE_Transparency} \\
Market prices & ENTSO-E transparency platform~\cite{ENTSOE_Transparency} \\
Photovoltaic (PV) production & Demand.ninja~\cite{staffell2025renewables} \\
Installed PV capacity & Swiss Federal Statistical Office~\cite{GWR} \\
Network data & Oneto et al.~\cite{Oneto2025a, Oneto2025b} \\
\hline
\end{tabular}
\label{tab:input_data}
\end{table}


\begin{table}[h!]
    \centering
    \caption{Overview of input parameters.}
    \renewcommand{\arraystretch}{1.1} 
    \begin{tabular}{llcc}
        \hline
        \textbf{Parameter} & \textbf{Description} & \textbf{Value} & \textbf{Unit} \\
        \hline
        \multicolumn{2}{l}{\textbf{Building}} & & \\
        \hline
        COP & Coefficient of performance & 4.0 & - \\
        $T^{\text{set}}$ & Set-point temperature  & 20.0 & \SI{}{\celsius}\\
        $T^{\text{min}}$ & Minimum indoor temperature  & 19.0 & \SI{}{\celsius}\\
        $T^{\text{max}}$ & Maximum indoor temperature  & 21.0 & \SI{}{\celsius}\\        
        \hline        
        \multicolumn{2}{l}{\textbf{Network}} & & \\
        \hline
        $V^{\text{min}}$ & Voltage lower bound (pu) & 0.97 & - \\
        $V^{\text{max}}$ & Voltage upper bound (pu) & 1.03 & - \\
        RAR & Reactive-to-active power ratio & 0.05 & - \\
        \hline
        \multicolumn{2}{l}{\textbf{Cost}} & & \\
        \hline
        $VoLL$ & Value of lost load & 10,000 & \euro{}/\SI{}{\mega\watt\hour} \\
        \hline
    \end{tabular}
    \label{tab:constants}
\end{table}

\subsection{Building model} \label{subsec: building_model}
\begin{figure}[ht]
\centering
\begin{tikzpicture}

  \draw[very thick] (0,0) rectangle (4,3);
  \draw[very thick] (0,3) -- (2,4.2) -- (4,3);

  \coordinate (Ti) at (2,1.5);
  \filldraw[black] (Ti) circle (2pt) node[above] {$T^\text{in}$};

  \draw (Ti) -- (2,0.9);
  \draw (1.7,0.9) -- (2.3,0.9);
  \draw (1.75,0.75) -- (2.25,0.75);
  \draw (2,0.75) -- (2,0.55); 
  \node at (2.7,0.8) {$C_{\text{}}$};

  \node at (-0.8,0.65) {\includegraphics[width=1.6cm]{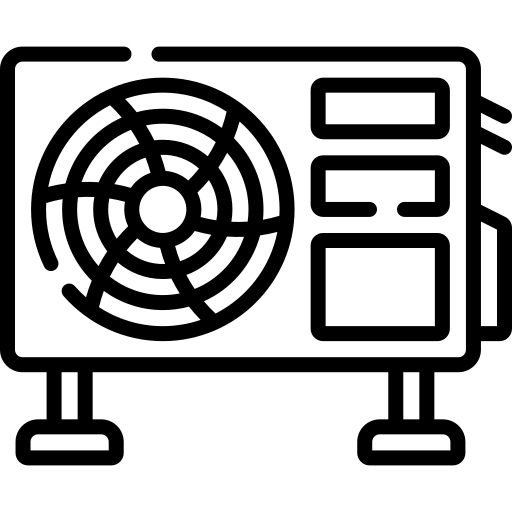}};

  \draw[->, thick] (-0.8,1.5) -- (1.7,1.5);
  \node at (0.8,1.8) {$\dot{Q}^\text{HP}$};

  \draw (Ti) -- (4.2,1.5);

  \draw (4.2,1.5) -- ++(0.2,0.3) -- ++(0.2,-0.6) -- ++(0.2,0.6)
                -- ++(0.2,-0.6) -- ++(0.2,0.3);  

  \draw (5.8,1.5) -- (5.2,1.5);

  \filldraw[black] (5.8,1.5) circle (2pt) node[above] {$T^{\text{out}}$};
  \node at (4.6,2.1) {$R_{\text{}}$};

  \draw[->, thick] (3.6,0.3) -- (5.2,0.3);
  \node at (4.6,0.6) {$\dot{Q}^{\text{loss}}$};

\end{tikzpicture}
\caption{Single-node RC building model.}
\label{fig: building model}
\end{figure}
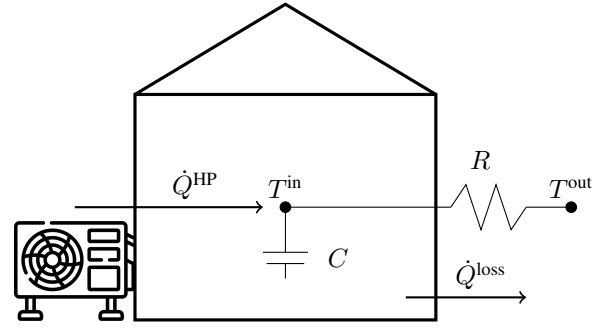

The load-shifting flexibility depends both on the technical constraints of the HPs and on the comfort requirements of the households involved. Therefore, the individual buildings are modeled as single-node resistance-capacitance (RC) equivalent circuits, as illustrated in Fig. \ref{fig: building model}. To quantify the value of flexibility, HPs are operated in either an inflexible or a flexible mode.

The temperature dynamics of each building are described by the following continuous energy balance:
\begin{equation*}
C \cdot \frac{dT^{\text{in}}}{dt} = \dot{Q}^{\text{HP}} - \dot{Q}^{\text{loss}}, 
\label{eq:temp_dynamics}
\end{equation*}
where $C$ is the building's thermal capacitance, $\dot{Q}^{\text{HP}}$ the heat supplied by the HP, and $\dot{Q}^{\text{loss}}$ the thermal losses through the building envelope. With buildings indexed by $b \in \mathcal{B} = \{1,\dots,B\}$, where $\mathcal{B}^{\text{HP}} \subseteq \mathcal{B}$ denotes the subset of buildings equipped with HPs, the discretized energy balance yields:
\begin{align}
T^{\text{in}}_{b,t} &= T^{\text{in}}_{b,t-1} 
    + \frac{\Delta t}{C_b} \cdot \Bigg(
        \text{COP} \cdot P^{\text{HP}}_{b,t} 
        - \frac{T^{\text{in}}_{b,t} - T^{\text{out}}_t}{R_b} 
    \Bigg), \nonumber \\
& \quad \forall b \in \mathcal{B}^{\text{HP}}, \; t \in \mathcal{T},
\label{eq:building_model}
\end{align}
where $R_b$ denotes the thermal resistance, $\text{COP}$ the coefficient of performance, $P^{\text{HP}}_{b,t}$ the electrical HP power, and $T^{\text{in}}_{b,t}$ and $T^{\text{out}}_t$ the indoor and outdoor temperatures, respectively. 

Inflexible HPs are assumed to maintain a constant set-point temperature $T^{\text{set}}$ by compensating thermal losses. 
The baseline HP power $P^{\text{HP,base}}_{b,t}$ is given by:
\begin{equation*}
P^{\text{base}}_{b,t} = \frac{T^{\text{set}} - T^{\text{out}}_t}{R_b \cdot\text{COP}}, \quad \forall b \in \mathcal{B}^{\text{HP}}, \; \forall t \in \mathcal{T}.
\label{eq:baseline_hp_power}
\end{equation*}

Flexible HPs are modeled as shiftable loads, whose daily energy consumption $E^{\text{base}}_b$ must equal the one for the inflexible case\footnote{We enforce equal energy consumption for flexible and inflexible HPs to isolate the effect of shifting consumption to lower price periods. Without this constraint, flexible HPs would tend to operate closer to the lower temperature bound, thereby consuming less independently of the shifting effect.}, and the instantaneous HP power $P^{\text{HP}}_{b,t}$ is only constrained by the rated HP power $P^{\text{HP,rated}}_{b}$:
\begin{equation*}
\begin{aligned}
0 \;\leq\; P^{\text{HP}}_{b,t} &\;\leq\; P^{\text{HP,rated}}_b, 
&& \forall b \in \mathcal{B}^{\text{HP}},\ \forall t \in \mathcal{T}, \\[8pt]
\sum_{t=1}^{T} P^{\text{base}}_{b,t} \cdot \Delta t 
&= E^{\text{base}}_b 
= \sum_{t=1}^{T} P^{\text{HP}}_{b,t} \cdot \Delta t, 
&& \forall b \in \mathcal{B}^{\text{HP}}.
\end{aligned}
\label{eq:hp_constraints}
\end{equation*}
Additionally, comfort is accounted for by imposing that the indoor temperature stays within predefined bounds:
\begin{equation*}
\begin{aligned}
T^{\text{min}} \;\leq\; T^{\text{in}}_{b,t} &\;\leq\; T^{\text{max}}, 
&& \forall b \in \mathcal{B}^{\text{HP}},\ \forall t \in \mathcal{T}, \\
T^{\text{in}}_{b,0} &= T^{\text{set}}, 
&& \forall b \in \mathcal{B}^{\text{HP}},
\end{aligned}
\label{eq:temperature_constraints}
\end{equation*}
where $T^{\text{min}}$ and $T^{\text{max}}$ denote the minimum and maximum temperature, respectively. The initial temperature $T^{\text{in}}_{b,0}$ is fixed to the set-point temperature $T^{\text{set}}$.

\subsection{Network model} \label{subsec: network model}
\begin{figure}[H]
    \centering
    \includegraphics[width=\columnwidth]{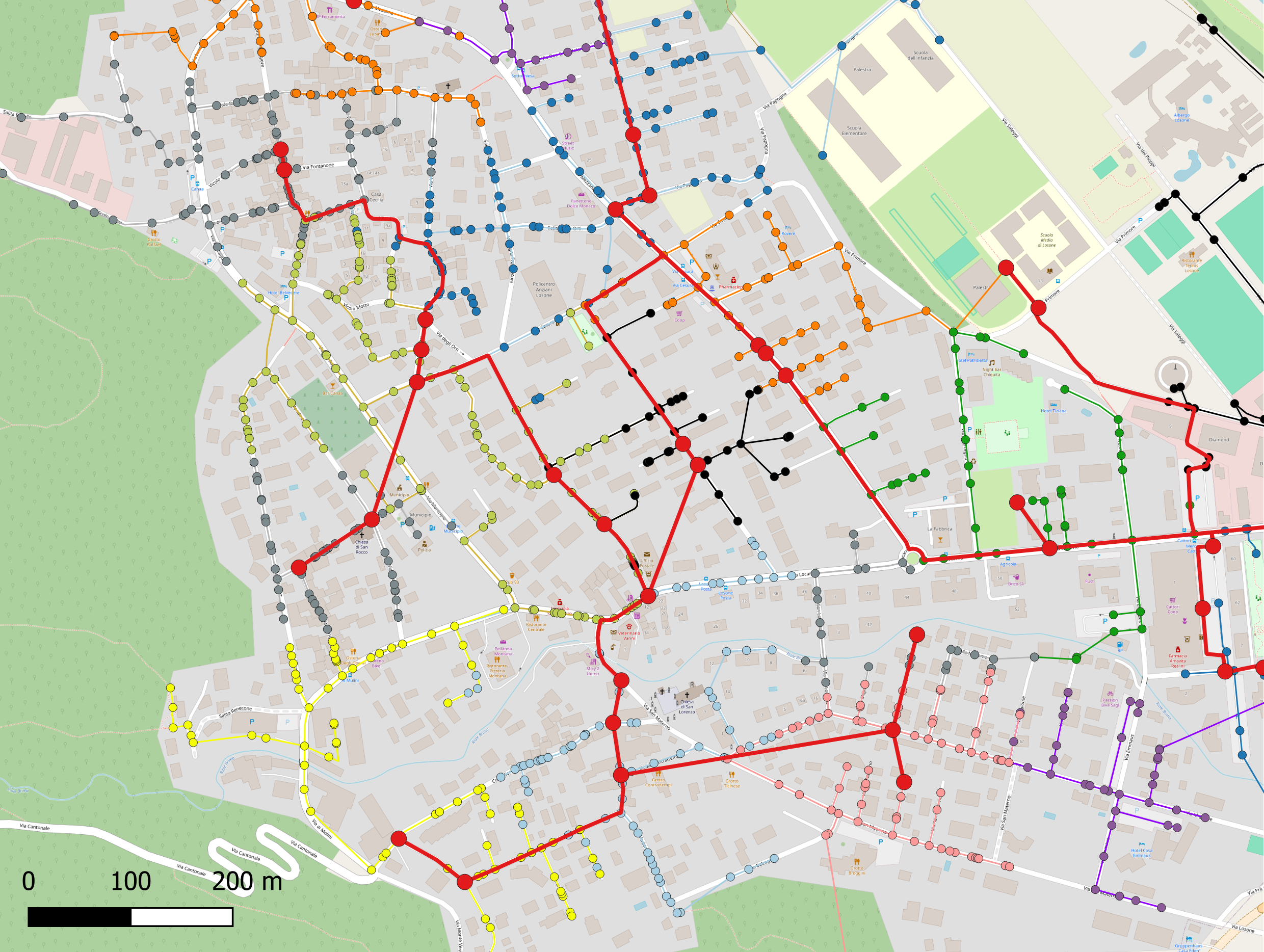}
    \caption[MV and LV distribution network of Losone]{Representation of the Losone MV and LV distribution network.}
    \label{fig:Losone_network}
\end{figure}

In Switzerland’s integrated retail market, the DSO is not only responsible for operating the distribution grid, but also supplying electricity to consumers. To represent this situation, the individual buildings are connected to the distribution grid of Losone. Since access to real-world electrical network data is restricted for privacy and security reasons, we use synthetic medium-voltage (MV) and low-voltage (LV) grid instances sourced from \cite{Oneto2025a}. These geo-referenced networks are derived from publicly available information and include node and edge data files with all necessary parameters \cite{Oneto2025b}. Fig. \ref{fig:Losone_network} illustrates the synthetic distribution network of Losone, with the MV network displayed in red and different LV networks distinguished by different colors. 

\paragraph{Building allocation}
Individual buildings with their respective HP and PV capacities are allocated to network nodes. This allocation is determined by solving optimization problem~\eqref{eq:allocation_problem}, which minimizes the distance between building and node positions $(X_b,Y_b)$ and $(X_n,Y_n)$, subject to assignment and capacity constraints. 
In this problem, each building $b \in \mathcal{B}$ is assigned to exactly one node $n \in \mathcal{N}$ through the binary variable $a_{n,b}$, while the node capacities $P^{\text{cap}}_n$ obtained from the synthetic dataset limit the rated HP power $P^{\text{HP,rated}}_b$ and PV power $P^{\text{PV,rated}}_b$.

\begin{equation}
\begin{aligned}
\min_{a_{n,b}} \quad & \sum_{n \in \mathcal{N}} \sum_{b \in \mathcal{B}} a_{n,b} \cdot \sqrt{(X_n - X_b)^2 + (Y_n - Y_b)^2} \\
\text{s.t.} \quad & \sum_{n \in \mathcal{N}} a_{n,b} = 1, \quad \forall b \in \mathcal{B} \\
& a_{n,b} \in \{0, 1\}, \quad \forall n \in \mathcal{N}, \; \forall b \in \mathcal{B} \\
& \sum_{b \in \mathcal{B}} a_{n,b} \cdot P^{\text{HP,rated}}_b \leq P^{\text{cap}}_n, \quad \forall n \in \mathcal{N}\\
& \sum_{b \in \mathcal{B}} a_{n,b} \cdot P^{\text{PV,rated}}_b \leq P^{\text{cap}}_n, \quad \forall n \in \mathcal{N}.
\end{aligned}
\label{eq:allocation_problem}
\end{equation}
After allocation, the aggregated HP and PV power at each node is obtained as the sum of the contributions from all connected buildings:
\begin{equation*}
\begin{aligned}
P^{\text{HP}}_{n,t} = \sum_{b \in \mathcal{B}^{\text{HP}} \cap \mathcal{B}^{n}} P^{\text{HP}}_{b,t}, \quad \forall n \in \mathcal{N}, \; \forall t \in \mathcal{T},\\
P^{\text{PV}}_{n,t} = \sum_{b \in \mathcal{B}^{\text{PV}} \cap \mathcal{B}^{n}} P^{\text{PV}}_{b,t}, \quad \forall n \in \mathcal{N}, \; \forall t \in \mathcal{T}.
\end{aligned}
\label{eq:hp_and_pv_aggregation}
\end{equation*}

\paragraph{Network constraints} 
When integrating the HPs into the distribution grid, their operation is constrained by the underlying network, governed by the power flow equations, and line, substations, and voltage limits must be satisfied. 
We assume radial distribution networks and use the linearized DistFlow formulation \cite{molzahn2019survey}. 
We adopt a sending-end convention for line flows where power is positive when directed upstream (towards the substation) and negative when directed downstream (away from the substation). 
Each node has a unique ancestor, namely its adjacent node in the upstream direction, and a set of child nodes, i.e., its adjacent nodes in the downstream direction. 
Nodal injections are defined as positive for generation and negative for consumption. 

The nodal power balances are described as follows:
\begin{equation*}
\begin{aligned}
P_{n,a,t} &= \sum_{c \in \mathcal{C}_n} P_{c,n,t} + P^{\text{inj}}_{n,t}, 
&& \forall n \in \mathcal{N},\ \forall t \in \mathcal{T},\\
Q_{n,a,t} &= \sum_{c \in \mathcal{C}_n} Q_{c,n,t} + Q^{\text{inj}}_{n,t}, 
&& \forall n \in \mathcal{N},\ \forall t \in \mathcal{T},
\end{aligned}
\label{eq:opf_balance}
\end{equation*}
where $P_{n,a,t}$ and $Q_{n,a,t}$ denote the active and reactive power flows from node $n$ to its ancestor $a$. The right-hand side consists of the sum of the flows $P_{c,n,t}, Q_{c,n,t}$ from all child nodes $c \in \mathcal{C}_n$ to node $n$, plus the net power injection $P^{\text{inj}}_{n,t}, Q^{\text{inj}}_{n,t}$ at node $n$.

Nodal injections are defined as follows: 
\begin{equation*}
\begin{aligned}
P^{\text{inj}}_{n,t} &= P^{\text{PV}}_{n,t} + P^{\text{LS}}_{n,t} - P^{\text{fix}}_{n,t} - P^{\text{HP}}_{n,t},
&& \forall n \in \mathcal{N},\ \forall t \in \mathcal{T},\\
Q^{\text{inj}}_{n,t} &= \text{RAR} \cdot \big(-P^{\text{fix}}_{n,t} - P^{\text{HP}}_{n,t}\big),
&& \forall n \in \mathcal{N},\ \forall t \in \mathcal{T},\\
\end{aligned}
\label{eq:opf_injection}
\end{equation*}
where $P^{\text{fix}}_{n,t}$ denotes the inelastic demand, $P^{\text{LS}}_{n,t}$ the load shedding (penalized by VoLL). The parameter RAR is a fixed reactive-to-active power ratio.

Line and substation loading limits are defined by
\begin{equation*}
\begin{aligned}
P_{l,t}^2 + Q_{l,t}^2 &\le S_l^2, 
&& \forall l \in \mathcal{L},\ \forall t \in \mathcal{T},\\
P_{n,t}^2 + Q_{n,t}^2 &\le S_n^2, 
&& \forall n \in \mathcal{N}^{sub} \subseteq \mathcal{N},\ \forall t \in \mathcal{T},
\end{aligned}
\label{eq:opf_limits}
\end{equation*}
where $S_l$ and $S_n$ are the apparent power ratings of line $l \in \mathcal{L}$ and substation node $n \in \mathcal{N}^{\text{sub}}$, respectively.

The voltage drop along each line $l = (i,j)$ connecting a sending node $i$ and a receiving node $j$ is given by 
\begin{equation*}
V_{i,t}^2 = V_{j,t}^2 - 2 \cdot (r_l \cdot P_{l,t} + x_l \cdot Q_{l,t}), \quad \forall l \in \mathcal{L}, \; \forall t \in \mathcal{T},
\label{eq:opf_voltage_drop}
\end{equation*}
where $r_l$ and $x_l$ are the resistance and reactance of line $l$, respectively. Voltage magnitude constraints are imposed by 
\begin{equation*}
\begin{aligned}
V^{\min} &\le V_{n,t} \le V^{\max}, 
&& \forall n \in \mathcal{N},\ \forall t \in \mathcal{T},\\
V_{n,t} &= V_{n}^{\text{nom}}, 
&& \forall n \in \mathcal{N}^{sub},\ \forall t \in \mathcal{T},
\end{aligned}
\label{eq:opf_voltage}
\end{equation*}
where $V^{\min}$ and $V^{\max}$ are the minimum and maximum voltage bounds, and $V_{n}^{\text{nom}}$ the fixed nominal voltage at all substation nodes.

The synthetic dataset does not provide any time series for the nodes but only the connection capacity. 
To obtain an hourly time series, we assume the same profile for each node.
That is, we compute the Load Factor for Switzerland, denoted by $\text{SLF}_t$ and defined as the ratio of the Swiss system load $P^{\text{CH}}_t$ at time $t$ to the annual peak load $P^{\text{CH,max}}$. 
The hourly electricity demand is obtained by scaling the capacity of the node by the SLF, i.e.,  $P^{\text{dem}}_{n,t} = P^{\text{cap}}_n \cdot \text{SLF}_t$.
The load constraints are then described by
\begin{equation*}
\begin{aligned}
P^{\text{fix}}_{n,t} &= P^{\text{dem}}_{n,t} - P^{\text{HP,base}}_{n,t}, 
&& \forall n \in \mathcal{N},\ \forall t \in \mathcal{T}, \\
0 \;\leq\; P^{\text{LS}}_{n,t} &\leq P^{\text{fix}}_{n,t}, 
&& \forall n \in \mathcal{N},\ \forall t \in \mathcal{T}.
\end{aligned}
\label{eq:load_constraints}
\end{equation*}
Hence, the inelastic demand is obtained by subtracting the baseline HP demand $P^{\text{HP,base}}_{n,t}$ from the total demand $P^{\text{dem}}_{n,t}$ and can be curtailed through load shedding.

The PV production is determined by
\begin{equation*}
\begin{aligned}
P^{\text{PV}}_{b,t} &= P^{\text{PV,rated}}_b \cdot \text{CF}_t, 
&& \forall b \in \mathcal{B}^{\text{PV}},\ \forall t \in \mathcal{T}, \\
\end{aligned}
\label{eq:pv_constraints}
\end{equation*}
where $\text{CF}_t$ denotes the capacity factor at time $t$.

\subsection{Bid determination} \label{subsec: results bid determination}
The bid determination consists of two major steps: the generation of different price scenarios and the determination of optimal power profiles based on these price scenarios. In our approach, aggregation reduces to a simple summation of the individual power profile vectors (cf. Algorithm~\ref{alg:agg_bid_selection}).

\paragraph{Price Scenarios} \label{paragraph: price forecasts}
Electricity prices exhibit pronounced daily, weekly and yearly patterns, enabling accurate prediction of future prices \cite{weron2014electricity}. For scenario generation, we use a two-step approach by first creating a point forecast and then applying a postprocessing method using the residuals from previous days to create the remaining scenarios. This approach is widely adopted in both literature and industry \cite{lohndorf2013optimizing} and has shown to perform competitively \cite{grothe2023point}.

In a first step, we use the open-source Python library \textit{epftoolbox} to create a point forecast $y_{d,t}$ using the \textit{LASSO Estimated Autoregressive} model \cite{lago2021forecasting}. The model is trained on historical Swiss DAM price data from 2020–2025 with a four-year calibration and a one-year test period. In a second step, $S$ price scenarios $\lambda_{d,t,S}$ are generated by incorporating residuals from the past $S-1$ days as follows:
\begin{equation*}
    \begin{aligned}
        \lambda_{d,t,1} &= y_{d,t}, \\
        \lambda_{d,t,2} &= y_{d,t} - (y_{d-1,t} - \lambda_{d-1,t}), \\
        &\;\;\vdots \\
        \lambda_{d,t,S} &= y_{d,t} - (y_{d-(S-1),t} - \lambda_{d-(S-1),t}),
    \end{aligned}
    \label{eq:price_scenario}
\end{equation*}
where the term $(y_{d-1,t} - \lambda_{d-1,t})$ represents the forecast residual from day $d-1$ and hour $t$. The computation time for generating the price scenarios is negligible.

\paragraph{Unbundled Utility}
For determining optimal power profiles, we adapt the generic formulation of the optimization problem~\eqref{eq: aggregation}. Since all power profiles accumulate to the same daily energy consumption $E^{\text{base}}$ and we assume that load shifting does not incur any cost, the valuation of all feasible power profiles $x_r \in \mathcal{X}_r$ are the same and proportional to the VoLL, given the utilities' obligation to supply the load:
\begin{equation*}
v_r(x_r) = \text{VoLL} \cdot \Delta t \cdot \sum_{t \in \mathcal{T}} x_{r}, 
\quad \forall x_r \in \mathcal{X}_r.
\end{equation*}
Because this valuation is constant across all feasible profiles, it can be omitted from the optimization problem, and \eqref{eq: aggregation} simplifies to a cost-minimization problem, which for an unbundled utility is described by
\begin{equation}
\min \sum_{t \in \mathcal{T}} \Delta t \cdot \Bigg( \lambda_{t} \cdot \sum_{b \in \mathcal{B}^{\text{HP}}} P^{\text{HP}}_{b,t} 
 \Bigg).
\label{eq:objective_function_unbundled}
\end{equation}
This optimization problem is solved repeatedly for all scenarios, and provides the power profiles for the block bids. In practice, we assume that the utility is willing to pay the maximum admissible bid price (MABP), currently at 4,000 \euro{}/\SI{}{\mega\watt\hour}~\cite{epexspot}. The resulting submitted block bids $\bar{x}_s$ and the scenario-independent bid price $p$ are then given by
\begin{equation*}
\bar{x}_s = \sum_{b \in \mathcal{B}^{\text{HP}}} P^{\text{HP}}_{b,s}, \qquad
p = \text{MABP} \cdot \sum_{b \in \mathcal{B}^{\text{HP}}} E^{\text{base}}_b.
\end{equation*}

\paragraph{Integrated Utility}
For an integrated utility, HP operation is further restricted by the constraints described in the network model in Section \ref{subsec: network model}. Therefore, the optimal dispatch problem extends to an OPF problem. As the share of HPs in the network increases (cf. Table~\ref{tab:hps_installed}), a point is reached where load shedding can no longer be avoided, since network reinforcements - which would occur in practice - are not modeled.
Consequently, the objective function~\eqref{eq:objective_function_bundled} must be augmented with a term that accounts for load shedding:
\begin{equation}
\min \sum_{t \in \mathcal{T}} \Delta t \cdot \Bigg( 
\lambda_{t} \cdot P^{\text{PCC}}_{t} 
+ VoLL \cdot \sum_{n \in \mathcal{N}} P^{\text{LS}}_{n,t} \Bigg),
\label{eq:objective_function_bundled}
\end{equation}
where $P^{\text{PCC}}_{t}$ denotes the power exchange at the point of common coupling with the high voltage grid, and $VoLL$ the value of lost load (or load shedding penalty). Since load shedding is highly penalized, flexibility is primarily leveraged for peak shaving, with the remaining flexibility used to react to price signals. 

\subsection{Benchmark - Aggregation Efficiency}

We compare our aggregation method with the theoretical optimum that could be achieved if it were possible to bid the entire aggregate feasible set $\mathcal{X}_{\mathrm{agg}}$. 
We define the aggregation efficiency $\eta$ as the realized share of the theoretical maximum cost savings:  
\begin{equation*}
\eta = \frac{TC^{\text{inf}} - TC^{\text{cleared}}}{TC^{\text{inf}} - TC^{\text{opt}}},
\label{eq:bid_savings}
\end{equation*}
where $TC^{\text{inf}}$ are the total costs under inflexible operation, $TC^{\text{cleared}}$ the total costs achieved by bidding an EG, and $TC^{\text{opt}}$ the minimum achievable total costs if $\mathcal{X}_{\mathrm{agg}}$ would have been submitted. The optimal cost $TC^{\text{opt}}$ corresponds to the case of perfect foresight of market prices. Hence, it is obtained by solving the optimization model with the realized market-clearing price $\tilde{\lambda}$ for the unbundled and integrated utility cases, as given in \eqref{eq:objective_function_unbundled} and \eqref{eq:objective_function_bundled}, respectively. This definition captures how effectively the aggregation and bidding approach translates into economic performance.

\subsection{Results} \label{subsec: results}
The performance of our aggregation approach is evaluated by two key performance indicators: the aggregation efficiency and the computation time. In particular, these indicators are analyzed with respect to the number of bids and the number of resources.

\paragraph{Aggregation Efficiency}
Fig.~\ref{fig:eff_bids} shows the aggregation efficiency for different numbers of submitted bids at a fixed 15\,\% HP share, with overlapping curves for the unbundled and integrated utilities. The case $B=1$ corresponds to submitting only a single schedule with a single block bid instead of using EGs. By employing EGs, aggregation efficiency can be increased from 91\,\% to 98\,\%, converging closely towards an optimal bid. Since these results were achieved with publicly available price forecasting tools, they can be seen as a lower bound for market agents with access to more advanced forecasting methodologies.
Our findings show that the aggregation efficiency increases monotonically with the number of submitted bids. Although the efficiency saturates, it remains advantageous to fully exploit the maximum allowed limit of 24 bids.

\begin{figure}[!htbp]
  \centering
  \begin{tikzpicture}[every node/.append style={font=\footnotesize,}]
    \begin{axis}[
      width=1.0\columnwidth,  
      height=0.625\columnwidth, 
      grid=both,  
      grid style={dashed, gray!30}, 
      xlabel={Number of Bids},
      ylabel={Aggregation Efficiency $\eta$ (\%)},
      xmin=0, xmax=25,      
      ymin=90, ymax=100,    
      ytick={90,92,94,96,98,100},
      legend style={at={(0.97,0.05)}, anchor=south east, font=\tiny, fill=none},
      legend columns=1
    ]

    \addplot [color=CBBlue, mark=o, thick] 
      table [x=NumberBids, y=System15, col sep=semicolon] {results/relative_savings_all.csv};
    \addlegendentry{Unbundled utility}

    \addplot [color=CBRed, mark=square, thick, dashed, mark options={solid}] 
      table [x=NumberBids, y=Combined15, col sep=semicolon] {results/relative_savings_all.csv};
    \addlegendentry{Integrated utility}

    \end{axis}
  \end{tikzpicture}
  \caption[Sampling efficiency for different numbers of submitted bids]{Aggregation efficiency for different numbers of submitted bids.}
  \label{fig:eff_bids}
\end{figure}
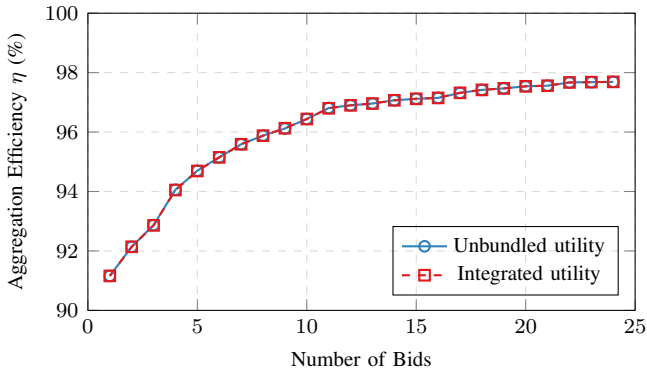

\begin{figure}[!htbp]
  \centering
  \begin{tikzpicture}[every node/.append style={font=\footnotesize,}]
    \begin{axis}[
      width=1.0\columnwidth,  
      height=0.625\columnwidth,       
      grid=both,  
      grid style={dashed, gray!30}, 
      xlabel={HP share (\%)},
      ylabel={Aggregation Efficiency $\eta$ (\%)},
      xmin=10, xmax=65,             
      xtick={15,30,45,60},             
      ymin=94, ymax=100,    
      ytick={90,92,94,96,98,100},
      legend style={at={(0.98,0.04)}, anchor=south east, font=\tiny, fill=none},
      legend columns=1
    ]



    \addplot [color=CBBlue, mark=o, thick] table [x=HPshare, y=24 bids, col sep=semicolon] {results/relative_savings_system_trans.csv};
    \addlegendentry{Unbundled utility}
    

    \addplot [color=CBRed, mark=square, thick, dashed, mark options={solid}] table [x=HPshare, y=24 bids, col sep=semicolon] {results/relative_savings_combined_trans.csv};
    \addlegendentry{Integrated utility}
    
    \end{axis}
  \end{tikzpicture}
  \caption[Sampling efficiency for different HP shares]{Aggregation efficiency for different HP shares.}
  \label{fig:eff_resources}
\end{figure}
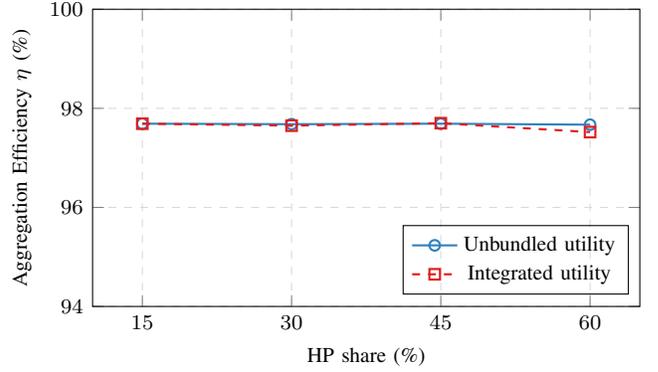

Fig.~\ref{fig:eff_resources} shows the aggregation efficiency for different numbers of resources, comparing the unbundled and integrated utility cases at a fixed bid number of 24. For the unbundled utility, the aggregation efficiency remains unaffected as the number of resources increases, highlighting that the proposed approach is scalable without efficiency loss. For the integrated utility, aggregation efficiency decreases slightly at higher HP shares due to arising grid congestion. 

\begin{figure}[!htbp]
  \centering
  \begin{tikzpicture}[every node/.append style={font=\footnotesize,}]
    \begin{axis}[
      width=1.0\columnwidth,  
      height=0.625\columnwidth,       
      grid=both,  
      grid style={dashed, gray!30}, 
      xlabel={HP share (\%)},
      ylabel={Computation Time (s)},
      xmin=10, xmax=65,             
      xtick={15,30,45,60},             
      ymin=0,
      ymax=400,    
      ytick={0,50,100,150,200,250,300,350,400},
      legend style={at={(0.03,0.97)}, anchor=north west, font=\tiny, fill=none},
      legend columns=1
    ]




    
    \addplot [color=CBBlue, mark=o, thick] table [x=HPshare, y=System24, col sep=semicolon] {results/comp_time_trans.csv};
    \addlegendentry{Unbundled utility}

    \addplot [color=CBRed, mark=square, thick, dashed, mark options={solid}] table [x=HPshare, y=Combined24, col sep=semicolon] {results/comp_time_trans.csv};
    \addlegendentry{Integrated utility}

    \end{axis}
  \end{tikzpicture}
  \caption{Computation time for different HP shares.}
  \label{fig:time_resources}
\end{figure}
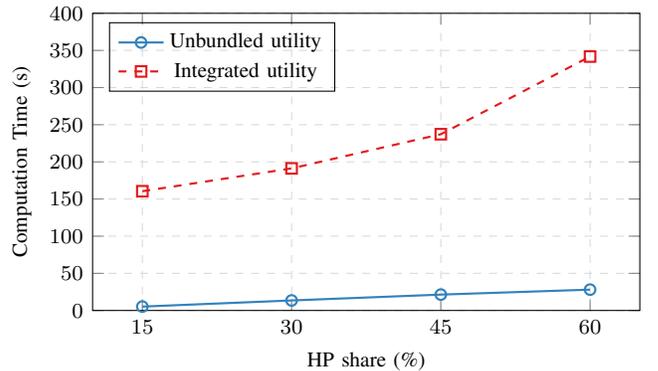

\paragraph{Computation time}
The cumulative computation time is illustrated for different numbers of bids at a fixed 45\,\% HP share in Fig.~\ref{fig:time_bids}, and for different HP shares at a fixed number of 24 bids in Fig.~\ref{fig:time_resources}, respectively. 

\begin{figure}[!htbp]
  \centering
  \begin{tikzpicture}[every node/.append style={font=\footnotesize,}]
    \begin{axis}[
      width=1.0\columnwidth,  
      height=0.625\columnwidth, 
      grid=both,  
      grid style={dashed, gray!30}, 
      xlabel={Number of Bids},
      ylabel={Computation Time (s)},
      xmin=0, xmax=25,      
      ymin=0, ymax=250,    
      ytick={0,50,100,150,200,250},
      legend style={at={(0.03,0.97)}, anchor=north west, font=\tiny, fill=none},
      legend columns=1
    ]

    \addplot [color=CBBlue, mark=o, thick] 
      table [x=NumberBids, y=System45, col sep=semicolon] {results/comp_time_all.csv};
    \addlegendentry{Unbundled utility}

    \addplot [color=CBRed, mark=square, thick, dashed, mark options={solid}] 
      table [x=NumberBids, y=Combined45, col sep=semicolon] {results/comp_time_all.csv};
    \addlegendentry{Integrated utility}

    \end{axis}
  \end{tikzpicture}
  \caption{Computation time for different bid numbers.}
  \label{fig:time_bids}
\end{figure}
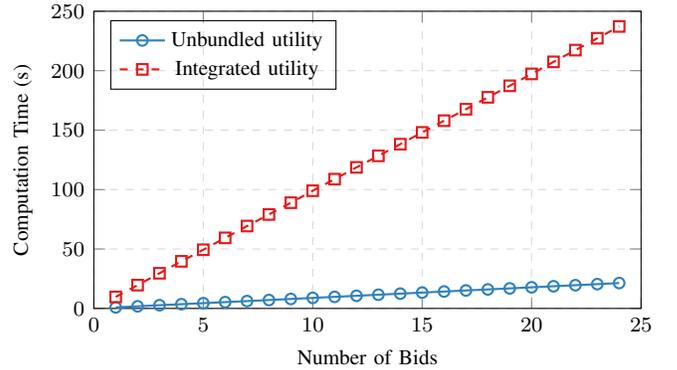

\begin{figure*}[t]
    \centering
    \includegraphics[width=\textwidth]{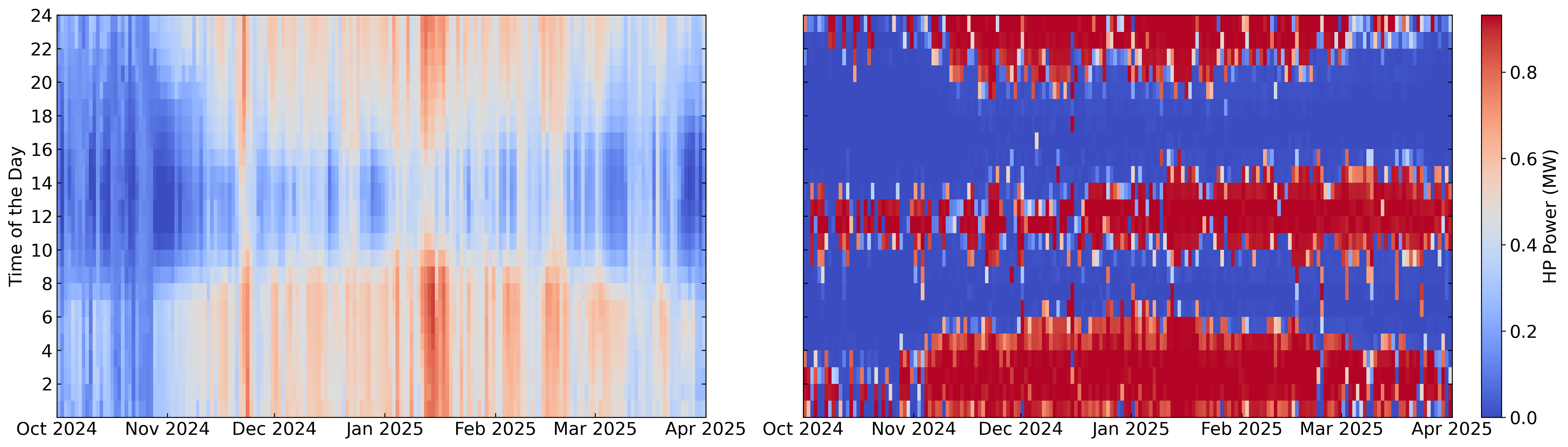}
    \caption[Comparison of consumption patterns for inflexible and flexible HPs]{Comparison of consumption patterns for inflexible (left) and flexible (right) HPs.}
    \label{fig:heat_map}
\end{figure*}

For the unbundled utility, computation time is directly proportional to the number of bids and resources. For the integrated utility, computation time is proportional to the number of bids, but shows a polynomial increase with the number of resources. This difference is due to the fact that the unbundled case decomposes into $R \cdot S$ optimal dispatch problems (cf. Algorithm~\ref{alg:agg_bid_selection}), while the integrated utility case requires solving $S$ OPF problems (cf. Section~\ref{paragraph: network}). In both cases, the proposed approach offers a computationally efficient method to determine bids within minutes.

\paragraph{Cost savings}
Taking advantage of the building inertia to respond to price signals alters the way in which HPs are operated. Fig.~\ref{fig:heat_map} illustrates the consumption profiles of inflexible and flexible HPs at a 15\,\% HP share. The operation of inflexible HPs follows outdoor temperature patterns, with slightly higher consumption during nighttime due to lower temperatures. In contrast, flexible HPs adjust their operation in response to price signals from the DAM. As a result, consumption is primarily shifted toward low-cost periods during nighttime and midday, and away from high-cost periods in the morning and evening. This shows that their operation aligns with system needs by reducing demand during peak hours and increasing it during times of high generation and low demand.

\begin{figure}
  \centering
  \begin{tikzpicture}[every node/.append style={font=\footnotesize,}]
    \begin{axis}[width=1.0\columnwidth, height=0.6\columnwidth, grid=both, grid style={dashed, gray!30}, xlabel={Standard Deviation of Hourly Prices within a Day (\euro{}/\SI{}{\mega\watt\hour})}, ylabel={Daily Cost Savings (\euro{}/HP)}, xmin=0, xmax=70, ymin=0, ymax=1.75, ytick={0,0.25,0.5,0.75,1.00,1.25,1.5,1.75}]

    \addplot[scatter,only marks,mark size=1.5pt,color=blue] table [x=StdDev,y expr=\thisrow{Total_Saving}/350,col sep=semicolon] {results/volatility.csv};
    \end{axis}
  \end{tikzpicture}
  \caption[Scatter plot of daily cost savings and price StdDev]{Scatter plot of daily cost savings and price volatility.}
  \label{fig:scatter_plot}
\end{figure}
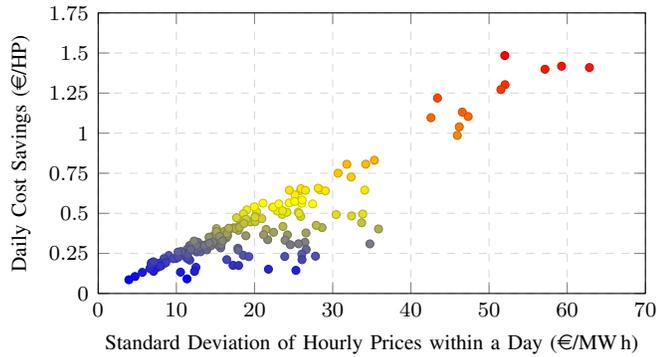

Our results highlight that the daily cost savings are closely related to the price volatility throughout the day. This is illustrated in Fig.~\ref{fig:scatter_plot}, where the relationship between daily cost savings and the standard deviation of hourly prices relative to the daily mean is shown for all days.

By taking advantage of volatile prices and occasional price dips, daily electricity costs can be reduced, leading to substantial overall savings. Over the considered time horizon, both the unbundled and integrated utility reduce electricity procurement costs by around 12.5\,\% across all HP shares compared to the inflexible operation, corresponding to cost savings of around 80~\euro{}/HP. Table \ref{tab:total_HP_costs} summarizes the total electricity procurement costs under perfect price foresight for different HP shares and operating modes.

For low HP shares, there is no noticeable difference in HP costs between the unbundled and integrated utility, indicating that the distribution grid is not constraining HP operation. For higher HP shares, grid-aware flexibility is prioritized, where load shifting is primarily used to perform peak shaving. 
\begin{table}[h!]
\centering
\caption{Total HP costs for the different operating modes.}
\renewcommand{\arraystretch}{1.1} 
\begin{tabular}{l c c c c}
\hline
Operating mode& 15\,\% HPs & 30\,\% HPs & 45\,\% HPs & 60\,\% HPs \\ \hline
Inflexible    & 212,891 \euro{} & 458,091 \euro{}& 685,263 \euro{}& 914,132 \euro{}\\
Unbundled   & 186,183 \euro{}& 400,655 \euro{}& 599,286 \euro{}& 799,682 \euro{}\\
Integrated & 186,183 \euro{}& 400,883 \euro{}& 599,706 \euro{}& 803,781 \euro{}\\ \hline
\end{tabular}
\label{tab:total_HP_costs}
\end{table}




\section{Conclusion} \label{sec: conclusion}
We proposed a method that enables an aggregator to bid aggregated flexibility in European day-ahead and intraday electricity auctions. To demonstrate the approach, we applied it to both an unbundled and an integrated utility in the town of Losone, where flexibility is provided by heat pumps and offered in the Swiss day-ahead auction. 

Our aggregation method offers the following advantages:
\begin{itemize}
    \item \textit{Technology-agnostic:} Compatible with all decentralized flexible resources (EVs, HPs, etc.).
    \item \textit{Market-ready:} Already applicable in most European day-ahead and intraday auctions.
    \item \textit{Scalable:} Computation time grows linearly with the number of resources when grid constraints are not considered.
    \item \textit{Accurate:} Delivers high accuracy given reasonably accurate electricity price forecasts.
\end{itemize}

The main limitation of our approach is its dependence on price forecasts. While forecast accuracy declines with higher renewable penetration, even imperfect forecasts provide useful signals for flexibility aggregation.

From our study, we can derive the following recommendations:
\begin{itemize}
    \item \textit{Aggregators} should utilize EGs to convey decentralized flexibility to the system.
    \item \textit{Nominated Electricity Market Operators (NEMOs)} who already offer EGs should increase the number of bids per group (e.g., from 24 to 100) to enable participants to express a larger share of their flexibility. NEMOs that are not yet supporting EGs should introduce them to facilitate the market integration of decentralized flexibility.
\end{itemize}




\bibliography{references}
%

\bibliographystyle{IEEEtran}


\newpage





\vfill

\end{document}